\documentclass[prl,twocolumn]{revtex4}
\usepackage{amssymb,amsmath,amsthm,graphicx}
\usepackage{latexsym}
\usepackage{bm}

\def\be{\begin{equation}}
\def\ee{\end{equation}}
\def\beq{\begin{eqnarray}}
\def\eeq{\end{eqnarray}}

\newcommand{\ie}{{\it i.e.,}\ }

\pdfoutput=1

\begin{document}

\def\lsim{\mathrel{\rlap{\lower4pt\hbox{\hskip1pt$\sim$}}
    \raise1pt\hbox{$<$}}}
\def\gsim{\mathrel{\rlap{\lower4pt\hbox{\hskip1pt$\sim$}}
    \raise1pt\hbox{$>$}}}
%\def\sqr#1#2{{\vcenter{\vbox{\hrule height.#2pt
%         \hbox{\vrule width.#2pt height#1pt \kern#1pt
%         \vrule width.#2pt}
%         \hrule height.#2pt}}}}
%\def\square{\mathchoice\sqr66\sqr66\sqr{2.1}3\sqr{1.5}3}
%%%%%%%%%%%%%%%%%%%%%%%%%%%%%%
\def\be{\begin{equation}}
\def\ee{\end{equation}}
\def\bea{\begin{eqnarray}}
\def\eea{\end{eqnarray}}
%%%%%%%%%%%%%%%%%%%%%%%%%%%%%%
\newcommand{\der}[2]{\frac{\partial{#1}}{\partial{#2}}}
\newcommand{\dder}[2]{\partial{}^2 #1 \over {\partial{#2}}^2}
\newcommand{\dderf}[3]{\partial{}^2 #1 \over {\partial{#2} \partial{#3}}}
\newcommand{\eq}[1]{Eq.~(\ref{eq:#1})}
\newcommand{\dd}{\mathrm{d}}

\title{Instability and new phases of higher-dimensional rotating black holes}

\author{\'Oscar J.~C.~Dias$^{a}$, Pau Figueras$^{b}$, Ricardo Monteiro$^{a}$, Jorge E.~Santos$^{a}$, Roberto Emparan$^{c}$\,}
\email{O.Dias@damtp.cam.ac.uk, pau.figueras@durham.ac.uk, R.J.F.Monteiro@damtp.cam.ac.uk, J.E.Santos@damtp.cam.ac.uk, emparan@ub.edu}
\affiliation{$\,$\\$^{a}$DAMTP, Centre for Mathematical Sciences,
    University of Cambridge, Wilberforce Road, Cambridge CB3 0WA, UK\\$\,$\\
$^{b}$Centre for Particle
Theory \& Department of Mathematical Sciences, Science Laboratories,
South Road, Durham DH1 3LE, UK\\$\,$\\
$^{c}$Instituci\'o Catalana de Recerca
i Estudis Avan\c{c}ats (ICREA), Passeig Llu\'{\i}s Companys 23,
E-08010 Barcelona, Spain, and\\
Departament de F\'{\i}sica Fonamental, Universitat de Barcelona, Marti i Franqu\`es 1,
E-08028 Barcelona, Spain}

\begin{abstract}
It has been conjectured that higher-dimensional rotating black holes
become unstable at a sufficiently large value of the rotation, and that
new black holes with pinched horizons appear at the threshold of the
instability. We search numerically, and find, the stationary
axisymmetric perturbations of Myers-Perry black holes with a single spin
that mark the onset of the instability and the appearance of the new
black hole phases. We
also find new ultraspinning Gregory-Laflamme instabilities of rotating
black strings and branes.
\end{abstract}

\maketitle

%%%%%%%%%%%%%%%%%%%%%%%%%%%%%%%%%%%%%%%%%%%%%%%%%%%%%%%%%%%%%%%%%%%%%%%%%%%
%%%%%%%%%%%%%%%%%%%%%%%%%%%%%%%%%%%%%%%%%%%%%%%%%%%%%%%%%%%%%%%%%%%%%%%%%%%

Black holes are the most basic and fascinating objects in General Relativity and the
study of their properties is essential for a better understanding of the
dynamics of spacetime at its most extreme. In higher-dimensional
spacetimes a vast landscape of novel black holes has
begun to be uncovered \cite{Emparan:2008eg}. Its layout --- \ie the
connections between different classes of black holes in the space of
solutions --- hinges crucially on the analysis of their classical
stability: most novel black hole phases are conjectured to branch-off at
the threshold of an instability of a known phase. Showing how this
happens is an outstanding open problem that we address in this paper.

\begin{figure}
\centering
\includegraphics[width = 6 cm]{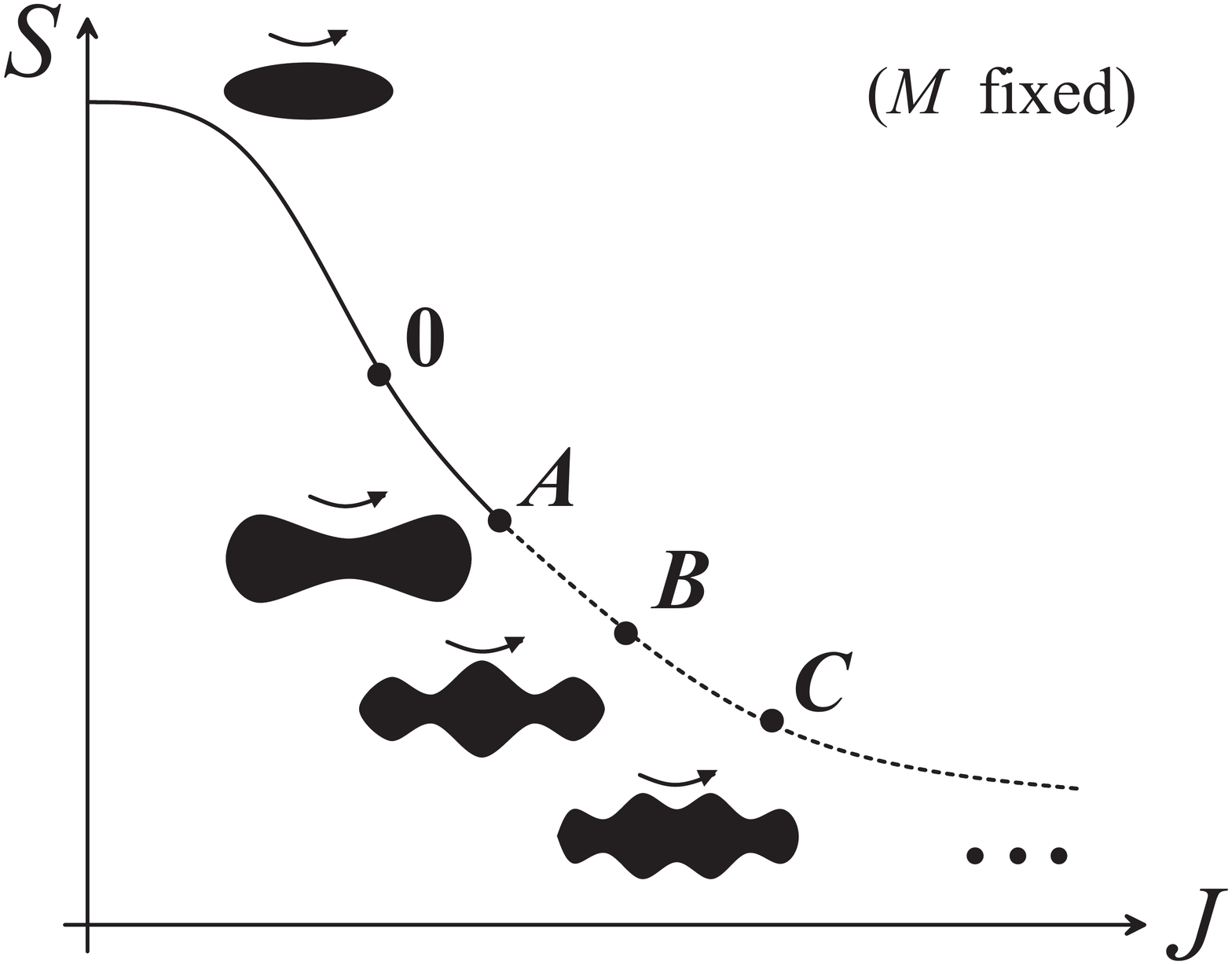}
\caption{\label{fig:phases} Diagram of entropy vs.\ spin, at fixed
mass, for MP black holes in $d\geq 6$ illustrating the conjecture of
\cite{Emparan:2003sy} (see also \cite{Emparan:2007wm}): at
sufficiently large spin the MP solution becomes unstable, and at the
threshold of the instability a new branch of black holes with a
central pinch appear ($A$). As the spin grows new of branches of
black holes with further axisymmetric pinches ($B,C,\dots$) appear.
We determine the points where the new branches appear, but it is not
yet known in which directions they run. We also indicate that at the
inflection point ($0$), where $\partial^2 S/\partial J^2=0$, there
is a stationary perturbation that should not correspond to an
instability nor a new branch but rather to a zero-mode that moves
the solution along the curve of MP black holes. }
\end{figure}

The best known class of higher-dimensional black holes, discovered by
Myers and Perry (MP) in \cite{myersperry}, appear in many respects as
natural generalizations of the Kerr solution. In particular, their
horizon is topologically spherical. However, the actual shape of the
horizon can differ markedly from the four-dimensional one, which is
always approximately round with a radius parametrically $\sim GM$. This
is not so in $d\geq 6$. Considering for simplicity the case where only
one spin $J$ is turned on (of the $\left\lfloor
\frac{d-1}{2}\right\rfloor$ independent angular momenta available), it
is possible to have black holes with arbitrarily large $J$ for a given
mass $M$. The horizon of these \textit{ultraspinning black holes}
spreads along the rotation plane out to a radius $a\sim J/M$ much larger
than the thickness transverse to this plane, $r_+\sim
(GM^3/J^2)^{1/(d-5)}$. This fact was picked out in \cite{Emparan:2003sy}
as an indication of an instability and a connection to novel black hole
phases. In more detail, in the limit $a\to\infty$ with $r_+$ fixed, the
geometry of the black hole in the region close to the rotation axis
approaches that of a black membrane. Black branes are known to exhibit
classical instabilities \cite{Gregory:1993vy}, at whose threshold a new
branch of black branes with inhomogeneous horizons appears
\cite{Gubser:2001ac}. Ref.~\cite{Emparan:2003sy} conjectured that this
same phenomenon should be present for MP black holes at finite but
sufficiently large rotation: they should become unstable beyond a
critical value of $a/r_+$, and the marginally stable solution should
admit a stationary, axisymmetric perturbation signalling a new branch of
black holes pinched along the
rotation axis. Simple estimates suggested that in fact
$(a/r_+)_\mathrm{crit}$ should not be much larger than one. As $a/r_+$
increases, the MP solutions should admit a sequence of stationary
perturbations, with pinches at finite latitude, giving rise to an
infinite sequence of branches of `pinched black holes' (see
fig.~\ref{fig:phases}). Ref.~\cite{Emparan:2007wm} argued that this
structure is indeed required in order to establish
connections between MP black holes and the black ring and black Saturn
solutions more recently discovered. Our main result is a numerical
analysis that proves correct the conjecture illustrated in
fig.~\ref{fig:phases}.

The solution for a MP black hole rotating in a single plane
in $d$ dimensions is \cite{myersperry}
\begin{eqnarray}\label{mpbh}
ds^2 &=& -dt^2 + \frac{r_m^{d-3}}{r^{d-5}\Sigma}\left( dt+a\sin^2\theta
\,d\phi\right)^2
+\Sigma\left(\frac{dr^2}{\Delta}+d\theta^2\right)\nonumber\\&&
+(r^2+a^2)\sin^2\theta\, d\phi^2 + r^2\cos^2\theta\, d\Omega^2_{(d-4)}\,,
\end{eqnarray}
\begin{equation}
 \Sigma=r^2+a^2\cos^2\theta\,,\qquad
\Delta=r^2+a^2-\frac{r_m^{d-3}}{r^{d-5}}\,.
\end{equation}
The parameters here are the mass-radius $r_m$ and the rotation-radius $a$,
\beq
r_m^{d-3}=\frac{16\pi
GM}{(d-2)\Omega_{d-2}}\,,\qquad
a=\frac{d-2}{2}\frac{J}{M}\,.
\eeq
The event horizon lies at the largest real root $r=r_+$ of $\Delta$.

The linearized perturbation theory of the Kerr black hole ($d=4$) was
disentangled in \cite{Teukolsky:1973ha} using the Newman-Penrose
formalism. Attempts to extend this formalism to decouple a master
equation for the gravitational perturbations of \eqref{mpbh} in $d\geq
5$ have failed so far. Moreover, even if some subsectors of the
perturbations of some classes of MP black holes have been decoupled
\cite{mpperts}, none of them shows signs of any instability and indeed
they do not contain the precise kind of perturbations we are interested
in. Thus we take a more frontal numerical approach to a full set of
coupled partial differential equations (PDE).

We intend to solve for a stationary linearized perturbation $h_{ab}$
around the background \eqref{mpbh}. Choosing traceless-transverse (TT)
gauge, $h^a_{\phantom{a}a}=0$ and $\nabla^a h_{ab}=0$, the equations to
solve are
\begin{equation} (\triangle_L h)_{ab}= - \nabla_c \nabla^c h_{ab} -2
R_{a\phantom{c}b}^{\phantom{a}c\phantom{b}d} h_{cd}=0\,,
\label{Lichnerowicz}
\end{equation}
where $\triangle_L$ is the Lichnerowicz operator in the TT gauge.
Actually, we solve the more general eigenvalue problem
\begin{equation} \label{eigenh}
(\triangle_L h)_{ab} =-k^2 h_{ab}\,,
\end{equation}
which is known to appear in two contexts: eqs.~\eqref{eigenh} determine the
stationary perturbations of a black
string in $d+1$ dimensions (obtained by adding a flat direction $z$ to
\eqref{mpbh}) with a profile $e^{ikz}h_{ab}$. Thus such modes with $k>0$
correspond to the threshold of the Gregory-Laflamme instability of black
strings \cite{Gregory:1993vy}. The same equations also describe
the negative modes of quadratic quantum corrections to the
gravitional Euclidean partition function \cite{Gross:1982cv}. A recent study of
this problem for the Kerr black hole has shown the existence of a branch
of solutions extending the negative Schwarzschild mode (with $k_\mathrm{Sch}\neq 0$) to
finite rotation, with $k$ growing as the rotation
increases towards the Kerr bound \cite{Monteiro:2009ke}.

Our reason to consider \eqref{eigenh} instead of trying to solve
directly for $k=0$ is that there exist powerful numerical methods for
eigenvalue problems that give the eigenvalues $k$ together with the
eigenvectors, \ie the metric perturbations. If the ultraspinning
instability is present for MP black holes in $d\geq 6$, then, in
addition to the analogue of the branch studied in
\cite{Monteiro:2009ke}, a new branch of negative modes extending to
$k=0$ must appear. The eigenvalue $k=0$ corresponds to a
(perturbative) stationary solution with a slightly deformed horizon.
In fact, as explained above, we expect an infinite sequence of such
branches that reach $k=0$ at increasing values of the rotation.
The solutions for $k>0$ imply new kinds of Gregory-Laflamme
instabilities and inhomogeneous phases of ultraspinning black strings
(see also \cite{Kleihaus:2007dg}).

The modes we seek preserve the $SO(d-3)\times SO(2)$ rotational
symmetries of the MP solution and depend only on the radial and polar
coordinates, $r$ and $\theta$ \cite{Emparan:2003sy}. Thus we take the
ansatz
\begin{eqnarray}
 ds^2&=&-e^{2\nu_0}(dt-\omega\, d\phi)^2+e^{2\nu_1}\,d\phi^2+
e^{2\eta}\sin^2\theta\, d\theta^2\nonumber \\
 &&+e^{2\gamma}\,(dr-\chi\,\sin\theta d\theta)^2
+e^{2\Phi}d\Omega^2_{d-4}\,.\label{ansatz}
\end{eqnarray}
We decompose a given quantity $Q=\{\nu_0,\nu_1,\eta,\gamma,\omega,\chi
\}$ as $Q=\overline{Q}+\delta Q$. The unperturbed contribution
$\overline{Q}(r,\theta)$ describes \eqref{mpbh}. The perturbations
$\delta Q(r, \theta)$ are determined solving the eigenvalue problem
\eqref{eigenh} subject to appropriate boundary conditions. After
imposing TT gauge, eq.~\eqref{eigenh} reduces to four coupled PDEs for
$\delta\eta$, $\delta \gamma$, $\delta\chi$ and $\delta\Phi$ (the TT
conditions then give $\delta\nu_0$, $\delta\nu_1$ and $\delta \omega$).
The boundary conditions are that the perturbations are regular and
finite at the horizon, $r=r_+$, at infinity, $r=\infty$, and at the
poles $\theta=0,\pi/2$. In addition, we impose $\delta\chi(r_+)=0$. It
is important to ensure that the eigenmodes we find are not pure gauge,
$h_{ab}=\nabla_{(a}\xi_{b)}$. We can prove that in the TT gauge, pure
gauge perturbations within our ansatz necessarily diverge at either the
horizon or infinity. Thus, with our boundary conditions, the eigenmodes
we obtain are never pure gauge.

We use a numerical approach
successfully applied to the identification of the negative
mode of Kerr and Kerr-AdS black holes \cite{Monteiro:2009ke}. It
employs a Chebyshev spectral numerical method (see
\cite{Monteiro:2009ke} for further details). We have carried out the
calculations for $d=7,8,9$. The cases $d=5$ (where the heuristics of
\cite{Emparan:2003sy} do not allow to predict any instability) and
$d=6$ present more difficult numerics. These, as well as a more detailed
presentation of our numerical approach, will be discussed elsewhere.

\begin{figure}
\centering
\includegraphics[width =5.0 cm]{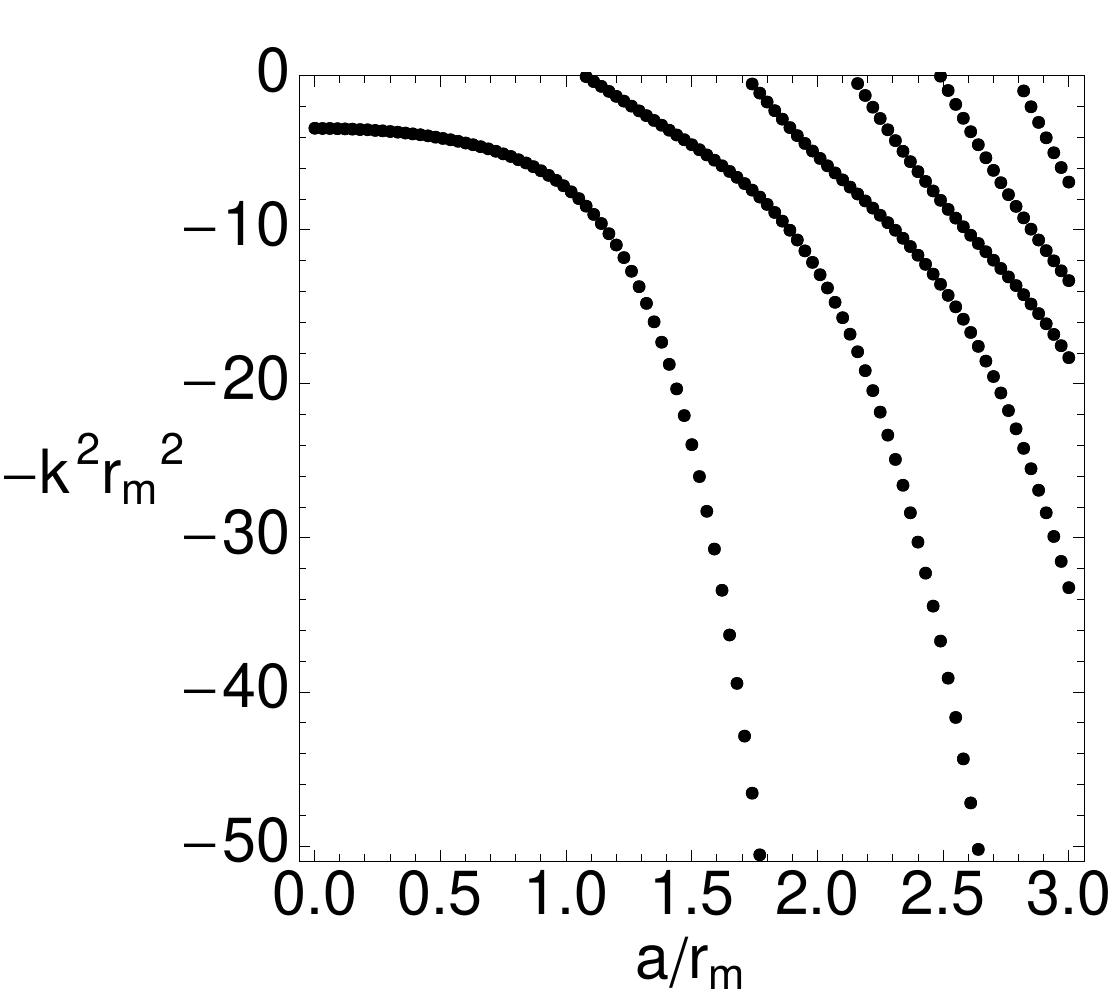}
\caption{\label{fig:negmode7d} Negative eigenvalues
for the MP black hole in $d=7$.}
\end{figure}
The results for $d=7$ are displayed in fig.~\ref{fig:negmode7d}, the
other two cases being qualitatively very similar. We plot the negative
eigenvalue $-k^2$ as a function of the rotation parameter $a$. We
normalize $k$ and $a$ relative to the mass-radius $r_m$, which is
equivalent to plotting their values for fixed mass (or mass per unit
length, in the black string interpretation). As described above, the
leftmost curve, which does not reach $k=0$, is the higher-dimensional
counterpart of the Kerr negative mode, and the eigenvalues $k$ are the
wavenumbers of the Gregory-Laflamme threshold modes at rotation $a$. At
larger rotation we find new branches of negative modes that intersect
$k=0$ at finite $a/r_m$. We label these successive branches with an
integer $\ell=1,2,3,\dots$, and refer to them as `harmonics'.
The values of $a/r_m$ at which the stationary perturbations appear are listed in
table~\ref{Table:critRot}.
%%%%%%%%%%%%%%%%%%%%
\begin{table}[ht]
\begin{eqnarray}
\nonumber
\begin{array}{|c|c|c|c|}\hline
 d & (a/r_m)|_{\ell=1} & (a/r_m)|_{\ell=2} &
 (a/r_m)|_{\ell=3} \\ \hline\hline
 7  & 1.075  & 1.714 & 2.141 \\
\hline
 8  & 1.061 & 1.770 & 2.275 \\
 \hline
9  & 1.051  & 1.792 & 2.337 \\
\hline
\end{array}
\end{eqnarray}
\caption{Values of the rotation $a/r_m$ for the first three harmonics of
stationary perturbation modes ($k=0$). The estimated numerical
error is $\pm 3\times 10^{-3}$ in $d=7$ and $\pm 5\times 10^{-3}$ in $d=8,9$.}
\label{Table:critRot}
\end{table}
%%%%%%%%%%%%%%%%%%%%%

It is important to note that the $k=0$ eigenmode of the harmonic
$\ell=1$ does {\it not} correspond to a new stationary solution.
Instead it is a  zero-mode that takes the solution to a nearby one
along the family of MP black holes. The existence and location of
this zero-mode is a consequence of the fact that if the Hessian of
the Gibbs potential $G(T,\Omega_i)=M-TS-\sum_i J_i \Omega_i$,
calculated along a family of solutions, has a zero eigenvalue for
some particular solution, then there is a  zero-mode perturbation of
the gravitational (Euclidean) action $I=G/T$ that takes that
solution to an infinitesimally nearby one along that family. That
is, perturbing the solution with that zero-mode does not correspond
to branching-off into a new family of solutions.

One can easily check that the determinant of the Hessian of the
Gibbs potential is proportional to the determinant of the Hessian of
the entropy with respect to only the angular momenta, \ie to the
determinant of
 \beq H_{ij}=\left(\frac{\partial^2 S}{\partial J_i
\partial J_j}\right)_M\,. \eeq

Therefore, for solutions with a single spin, there must appear a
stationary perturbation, in principle not associated to an
instability of the black hole, at the inflection point of the curve
$S(J)$ at fixed $M$ (point $0$ in fig. \ref{fig:phases}).
 For the MP solutions this happens at
\beq
\label{EMestimate}
 \left(\frac{a}{r_m}\right)^{d-3}_\mathrm{mem}=
\frac{d-3}{2(d-4)}\left(\frac{d-3}{d-5}\right)^{\frac{d-5}{2}}\,.
\eeq
The values of $(a/r_m)_\mathrm{mem}$ for $d=7,8,9$ agree
with the central values of the numerically-determined rotations
$(a/r_m)$ for $\ell=1$ (first column in table~\ref{Table:critRot}) up to the
third decimal place. This is a very good check of the accuracy of our
numerical methods.

The $k=0$ eigenmodes of the higher harmonics, $\ell\geq 2$, do not admit
this interpretation as perturbations along the MP family of solutions
and thus correspond to genuinely new (perturbative) black hole solutions
with deformed horizons. Their appearance conforms perfectly to the
predictions in \cite{Emparan:2003sy} and \cite{Emparan:2007wm}. It is
then natural to expect, although our approach does not prove it since it
only captures zero-frequency perturbations, that the
harmonic $\ell=2$ signals the onset of the instability conjectured in
\cite{Emparan:2003sy}. The $k=0$ eigenmodes for higher harmonics confirm
the appearance of the sequence of new black hole phases as the rotation
grows.

\begin{figure}
\centering
\includegraphics[width = 5 cm]{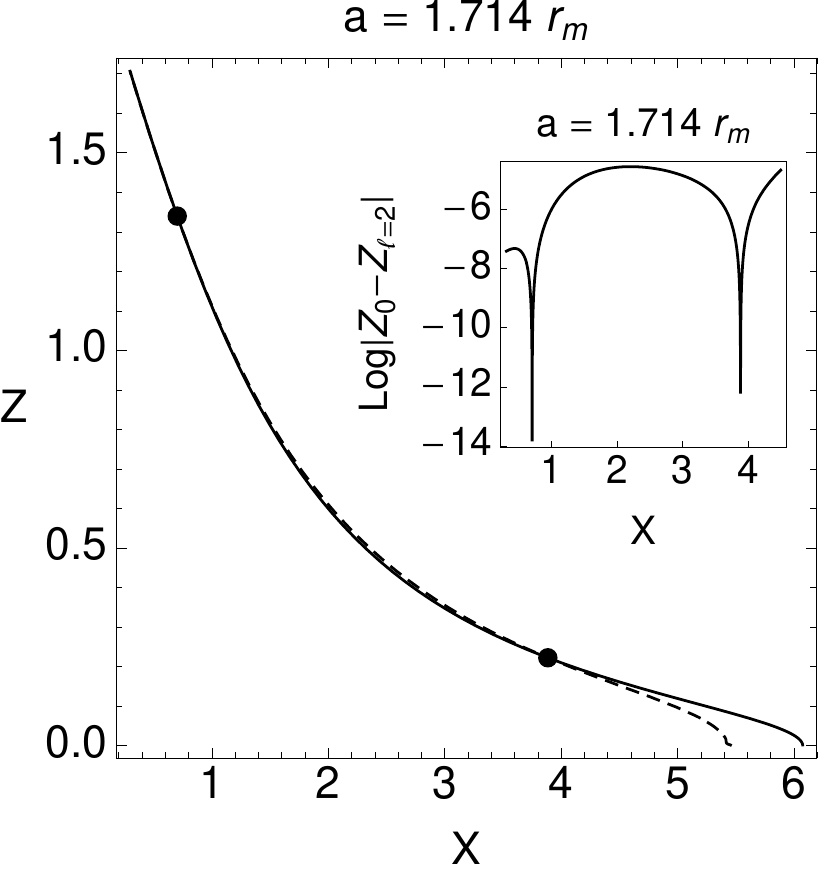}
\caption{\label{fig:embeddingl2} Embedding diagram at
$(a/r_m)_\mathrm{crit}$ of the $d=7$ black hole horizon, unperturbed
(solid), and with the first unstable harmonic perturbation
($\ell=2$, $k=0$) (dashed). The embedding Cartesian coordinates $Z$
and $X$ lie resp.\ along the rotation axis $\theta=0$ and the
rotation plane $\theta=\pi/2$. We also show the logarithmic
difference between the embeddings of the perturbed ($Z_{\ell=2}$)
and unperturbed ($Z_0$) horizons. The spikes represent the points
where the two embeddings intersect. The perturbation has two nodes,
so the horizon squeezes around the rotation axis, then bulges out,
and squeezes again at the equator, as in the conjectured shape $A$
in fig.~\ref{fig:phases}. }
\end{figure}

\begin{figure}
\centering
\includegraphics[width = 5 cm]{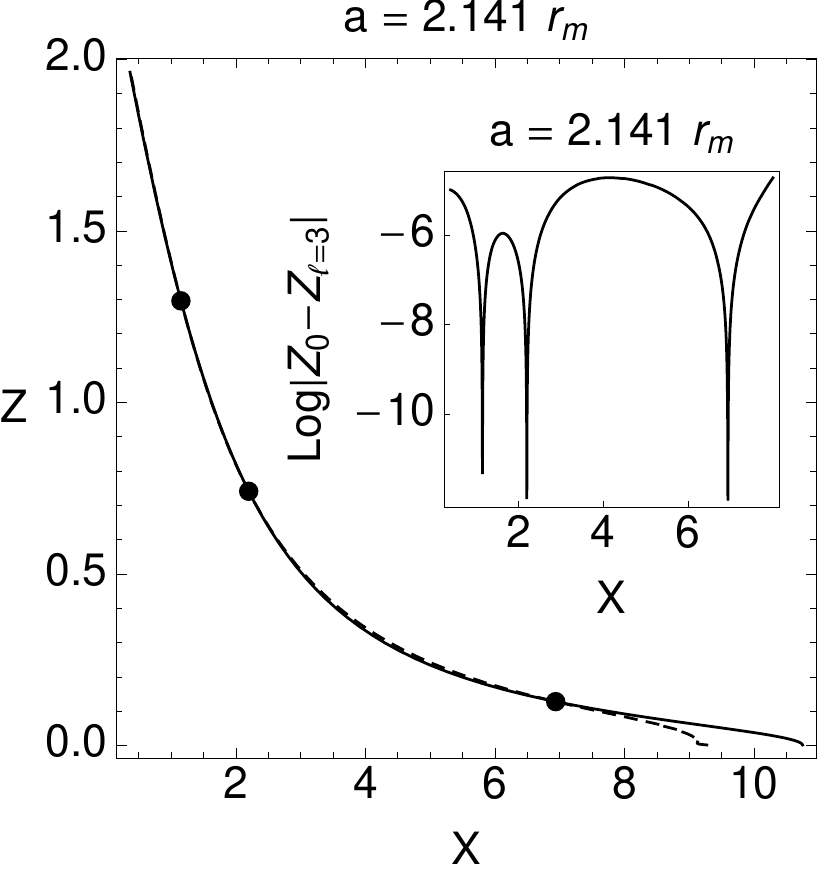}
\caption{\label{fig:embeddingl3} Like fig.~\ref{fig:embeddingl2},
for $\ell=3$: between the first two nodes of the perturbation the
horizon has a pinch (shape $B$ in fig.~\ref{fig:phases}).}
\end{figure}

\begin{figure}
\centering
\includegraphics[width = 4.8 cm]{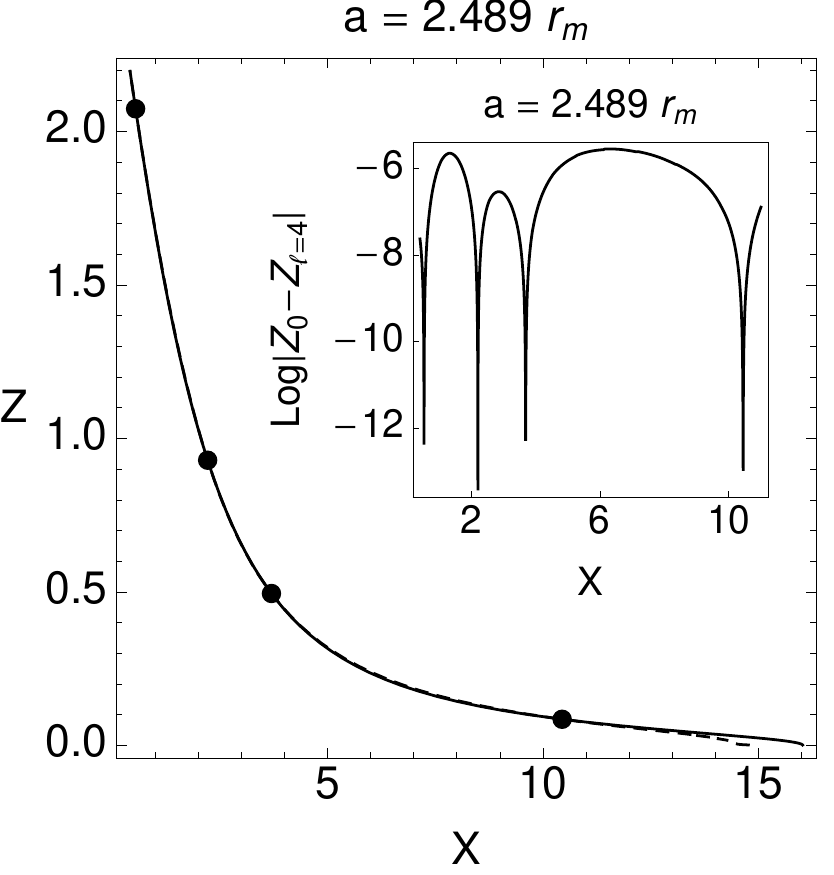}
\caption{\label{fig:embeddingl4} Like fig.~\ref{fig:embeddingl2},
for $\ell=4$: the four nodes deform the horizon into shape $C$ of
fig.~\ref{fig:phases}.}
\end{figure}

To visualize the effect on the horizon of the perturbations that
give new solutions, and provide further confirmation of our
interpretation, we draw an embedding diagram of the unperturbed MP
horizon and compare it with the deformations induced by the
ultraspinning harmonics $\ell\geq 2$. This is best done using the
embedding proposed in \cite{Frolov:2006yb}, which has the advantage
of allowing to embed the horizon along the entire range $0\leq
\theta\leq \pi/2$ for any rotation, although at the cost of
stretching the pole region, which acquires a conical profile. We do
it for the $\ell=2,3,4$ ultraspinning harmonics in
figs.~\ref{fig:embeddingl2}--\ref{fig:embeddingl4}. In spite of the
distortion created by the embedding, the effect of the perturbations
is clear: $\ell=2$ modes create a pinch centered on the rotation
axis $\theta=0$; $\ell=3$ modes have a pinch centered at finite
latitude $\theta$; $\ell=4$ modes pinch the horizon twice: around
the rotation axis and at finite latitude. These are the kind of
deformations depicted in fig.~\ref{fig:phases}. To better identify
the number of times that the perturbed horizon crosses the
unperturbed solution, in these figures we also plot the logarithmic
difference between the two embeddings.


Ref.~\cite{Emparan:2003sy} gave several arguments to the effect that
critical values $a/r_m$ close to 1 were to be expected. In
particular, it was pointed out that the change in the behavior of
the black hole from `Kerr-like' to `black membrane-like' could be
pinpointed to the value of the spin where the temperature (\ie
surface gravity) reaches a minimum for fixed mass, which is the
same, for solutions with a single spin, as the inflection point of
$S(J)$. As we have argued, the zero-mode at this solution should not
signal an instability. The $\ell=2$ mode at the threshold of the
actual instability instead appears at larger rotation, well within
the regime of membrane-like behavior as conjectured in
\cite{Emparan:2007wm}. We expect this to be true in general: the
ultraspinning instability of MP black holes should appear for
angular momenta {\it strictly beyond} the (codimension 1) locus in
the space of angular momenta where the Hessian $H_{ij}$ has a zero
eigenvalue.

In particular, in $d=5$ this criterion does not allow any ultraspinning
instability for any $J_1$, $J_2$, and in $d\geq6$ with all the
$N=\left\lfloor \frac{d-1}{2}\right\rfloor$ angular momenta $J_i$ equal
it predicts that the instability should appear at $a/r_m>2^{-N/(d-3)}$.
However we cannot predict the precise values of the rotation where the
instability appears.

We have identified the points in the phase diagram where the new
branches must appear, but we cannot determine in which direction these run.
This requires calculating the area, mass and spin of the perturbed
solutions. However, for any $k\neq 0$ --- and numerically we can
never obtain an exact zero --- the linear perturbations decay
exponentially in the radial direction, and so the mass and spin,
measured at asymptotic infinity, are not corrected. It seems that in
order to obtain the directions of the new branches one has to go beyond
our level of approximation or adopt a different approach.

The new $\ell\geq 1$ branches extend to non-zero eigenvalues $k$. These
imply a new ultraspinning Gregory-Laflamme
instability for black strings, in which the horizon is deformed
not only along the direction of the string, but
also along the polar direction of the transverse sphere. Observe that,
even if the $\ell=1$, $k=0$ mode is not an instability of the MP black
hole, the modes $\ell=1$, $k>0$ are expected to correspond to thresholds
of GL instabilities of MP black strings.
At
a given rotation, modes with larger $\ell$ have longer wavelength
$k^{-1}$ and so the branch $\ell=1$ is expected to dominate the
instability. The growth of $k$ with $a$ can be understood heuristically,
since as $a$ grows the horizon becomes thinner in directions transverse
to the rotation plane and hence it can fit into a shorter compact
circle.

To finish, we mention that pinched phases of rotating plasma balls, dual
to pinched black holes in Scherk-Schwarz compactifications of
AdS, have been found \cite{Lahiri:2007ae}, as well as new kinds of
deformations of rotating plasma tubes \cite{Caldarelli:2008mv} and
rotating plasma ball instabilities \cite{Cardoso:2009bv}. The
relation of our results to these and other phenomena of rotating fluids
will be discussed elsewhere.

\smallskip

%%%%%%%%%%%%%%%%%%%%%%%%%%%%%%%%%%%%%%%%%%%%%%%%%%%%%%%%%%%%%%%%%%%%%%%%%%
\noindent{\bf Acknowledgments.}
We thank Troels Harmark, Keiju Murata, Malcolm Perry and
especially Harvey Reall for discussions. We were supported by:
Marie Curie contract PIEF-GA-2008-220197, and by PTDC/FIS/64175/2006,
CERN/FP/83508/2008~(OJCD); STFC Rolling grant~(PF); Funda\c c\~ao para
a Ci\^encia e Tecnologia (Portugal) grants
SFRH/BD/22211/2005~(RM), SFRH/BD/22058/2005~(JES); and by
MEC~FPA~2007-66665-C02 and CPAN~CSD2007-00042
Consolider-Ingenio~2010~(RE). This is preprint
DCPT-09/47.

%%%%%%%%%%%%%%%%%%%%%%%%%%%%%%%%%%%%%%%%%%%%%%%

\end{document}